# From cluster to nanocrystal: the continuous evolution and critical size of copper clusters revealed by machine learning

Hongsheng Liu[1], Luneng Zhao[1], Yaning Li[1], Yuan Chang[1], Shi Qiu[1], Xiao Wang[2,3], Junfeng Gao[1,2]*, Feng Ding[2]

1. Key Laboratory of Materials Modification by Laser, Ion and Electron Beams (Dalian University of Technology), Ministry of Education, School of Physics, Dalian 116024, China
2. Suzhou Laboratory, Suzhou 215123, China
3. Shenzhen Institutes of Advanced Technology, Chinese Academy of Sciences

Corresponding author: (gaojf@dlut.edu.cn)

**Abstract**

The evolution of cluster structure with size and the critical size for the transition from cluster to nanocrystal have long been fundamental problems in nanoscience. Due to limitations of experimental technology and computational methods, the exploration of the continuous evolution of clusters towards nanocrystal is still a big challenge. Here, we proposed a machine learning force field (MLFF) that can generalize well to various copper systems ranging from small clusters to large clusters and bulk. The continuous evolution of copper clusters $Cu_N$ towards nanocrystal was revealed by investigating clusters in a wide size range ($7 \leq N \leq 17885$) based on MLFF simulated annealing. For small $Cu_N$ ($N < 40$), electron counting rule plays a major role in stability. For large $Cu_N$ ($N > 80$), geometric magic number rule plays a dominant role and the evolution of clusters is based on the formation of more and more icosahedral shells. For medium size $Cu_N$ ($40 \leq N \leq 80$), both rules contribute. The critical size from cluster to nanocrystal was calculated to be around 8000 atoms (about 6 nm in diameter). Our work terminates the long-term challenge in nanoscience, and lay the methodological foundation for subsequent research on other cluster systems.

## 1. Introduction

Clusters, a state of matter intermediate between individual atoms/molecules and the solid state, exhibit a rich variety of physical and chemical properties, which change significantly with cluster size.[1, 2] Therefore, clusters have great application prospects in many fields, including chemical catalysis[3-8], energy storage[9-11], optics[12-17], and biomedicine[18-21]. Moreover, clusters are building blocks for atomic manufacturing. Cu clusters have attracted much attention because of the high yield of Cu under mild synthesis conditions, as well as the abundance and low cost of Cu, which offers practical possibilities for large-scale nanotechnology applications, including catalysis, biomedicine, sensing and optoelectronics[22-35].

The structure determines the properties of a cluster, which are crucial for applications. Therefore, the evolution of cluster structure with size has long been a fundamental problem in nanoscience. However, the determination of cluster structure has always been a challenging subject. Up to now, it is impossible to obtain clear atomic resolution images of clusters in experiments and experimental spectra cannot directly determine the structure of clusters. Theoretical calculation is expected to play an important role in finding the ground state structure of a cluster.

Over the past few decades, a great deal of effort has been devoted to searching the ground-state structures of copper clusters[36-51]. At an earlier time, empirical potentials were used to study the ground state structures of clusters and subsequently tight-binding models were employed to study this problem. For example, Mukul Kabir et al. investigated the structure and stability of $Cu_N$ (N=3-55) clusters using a tight-binding molecular dynamics method[37]. However, the accuracy of empirical potentials and semi-empirical methods was unsatisfactory. Therefore, more precise quantization calculations, such as the DFT, were later used to study the ground-state structure of clusters. For instance, Galip H. Guvelioglu et al. systematically studied the structural evolution of small copper clusters with 2-15 atoms using DFT[39]. Calaminici et al. explored the lowest-energy isomers of $Cu_N$ (N = 12, 14, 16, 18, 20) clusters [52]. The ground-state structure of $Cu_{55}$, which is considered to have high stability, was carefully studied using DFT[47]. However, these works only sporadically studied several small-sized copper clusters, and there is still a lack of understanding of the overall structural evolution of copper clusters. This is because as the number of atoms in the cluster increases, the number of isomers increases exponentially and the potential energy surface becomes extremely complex. The DFT has high computational accuracy, but its computational cost is enormous. The time and spatial scales of the systems simulated by the DFT are severely limited. On the other hand, while traditional empirical potentials can be simulated on larger time and space scales, their accuracy is relatively low. Limited by simulation methods, the investigation of the evolution of cluster structure with size is still a challenge.

In recent years, the development of MLFF has provided new insights into addressing the challenge of balancing accuracy and simulation scale in theoretical simulations.[53] Wang et al. proposed a program for quickly predicting low-energy structures of nanoclusters by combining genetic algorithms with interatomic potentials actively learned on-the-fly and used it to search for $Al_N$ (N=20-55) cluster structures[54]. Fang et al. adopted and utilized a machine learning method based on kernel rigid regression to accelerate the search for low-energy structures of ligand protected clusters[55]. Yang et al. proposed a new method combining deep neural networks and transfer learning to improve the performance of global optimization of Pt metal cluster structures, significantly reducing the required number of samples and saving computational time[56]. By combining global optimization and the deep learning model, Wang et al. investigated a series of Cu clusters, $Cu_N$ (N = 10 – 50) and proposed that small clusters (N = 10 – 15) tends to be oblate and it gradually transforms into a cage-like configuration as the size increases (N > 15).[57] However, these studies only focus on clusters within a specific size range. Research into the evolution of cluster structures, especially the critical size at which clusters transition to bulk structures, requires a MLFF applicable to all sizes of clusters.

In this paper, by rationally constructing the dataset from DFT calculations, a

MLFF that can well describe Cu systems covering clusters, surfaces and crystals was proposed based on a local equivariant deep neural network. Combining this MLFF and simulated annealing methods, the low energy structures of copper clusters $Cu_N$ in a large size range ($7 \leq N \leq 17885$) were investigated. Our results show that Cu atoms in clusters prefer to be 12-coorninated but the packing mode is based on icosahedrons which is different from face centered cubic (FCC) close packing in the crystal. The critical size at which cluster structure evolves into crystal structure is calculated to be around 8000 atoms. Our work paves the way for subsequent studies on the structural evolution of other clusters.

## 2. Results and Discussion

### 2.1 Training and performance of the machine learning force fields

To obtain a universal MLFF suitable for copper clusters of different sizes, our dataset contains $Cu_N$ clusters with N = 12, 13, 14, 54, 55, 56, 146, 147, 148. For each size of copper clusters, at least 10 different structures were constructed followed by ab initio molecular dynamics simulations. Moreover, DFT calculations for Cu dimer, Cu bulk and Cu surfaces ((001, (110) and (111) surfaces) were also included in the dataset to enhance the generalization ability of the MLFF. The structure of the dataset was visualized using principal component analysis (PCA), which represents the latent space in two dimensions (Figure 1a). The latent space was not clearly distinguished for each composition, but different structures can be distinguished by their formation energy. The formation energy of bulks is lower, while the formation energy of small-sized clusters is higher. For the training, 4000 (80%) data was used as the training set and 1000 (20%) data as the validation set. As shown in Figure 1b-c, the trained MLFF performs well on the validation data set, with a good agreement between DFT and MLFF for both energies and forces. For energy, the root mean squared error (RMSE) and mean absolute error (MAE) are 4.2 meV/atom and 2.9 meV/atom, respectively. For the force, the RMSE and MAE are 26.0 meV/Å and 19.1 meV/Å, respectively.

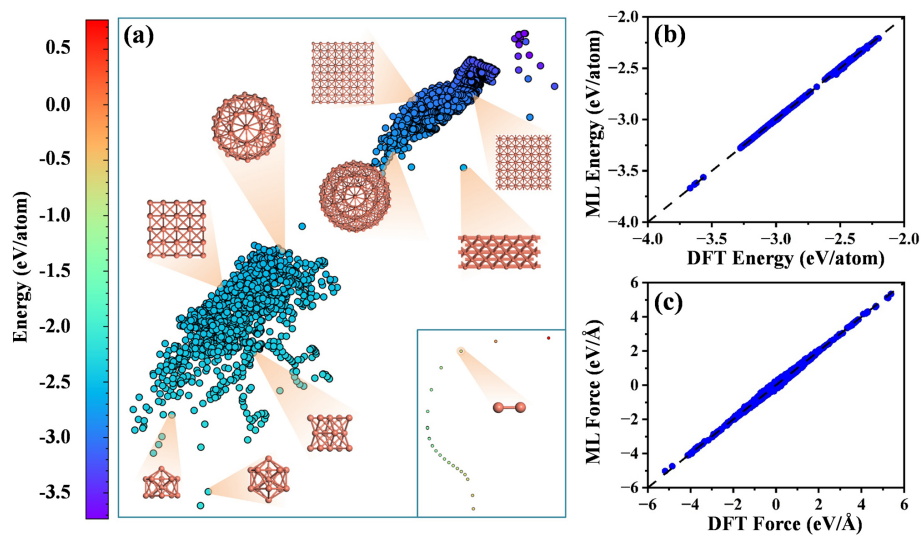

**Figure 1.** (a) Visualization of latent space through principal component analysis. The structures can be represented as circles of different colors based on its formation energy. (b) Energy comparison between machine learning (ML) and DFT for the validation data set. (c) Force comparison between ML and DFT for the validation data set. The dashed lines in (b) and (c) indicate the perfect agreement between ML and DFT.

To prove the accuracy and generalization ability of the MLFF, copper clusters $Cu_N$ of different sizes (N = 14, 24, 34, ..., 254) that are not included in the dataset were constructed. The energies of these clusters calculated by DFT and MLFF were compared as shown in Figure 2a. Most of the points fall on the ideal bisector, indicating the good generalization ability of the MLFF. To further verify that our MLFF can accurately perform the simulated annealing process, we randomly selected several structures from simulated annealing of clusters in different sizes, copper bulk, and surfaces, and calculated their energies using DFT and MLFF. The comparison of the DFT and MLFF energies together with the annealing process are shown in the Figure 2b-f. Figure 2b-d shows the annealing processes for $Cu_{41}$, $Cu_{55}$ and $Cu_{147}$ to get the ground state structures. Figure 2e shows the annealing process for a supercell of a face centered cubic copper crystal and Figure 2f shows the annealing process for three-layer surface model of Cu crystal. For all these annealing processes, the MLFF's description of the energy sequence of different structures during annealing is completely consistent with the DFT results. Therefore, our MLFF has very good universality and can describe various copper systems including copper clusters of different sizes, copper crystal and copper surfaces. Moreover, the computational speed of the MLFF is significantly faster than DFT, and the difference in speed becomes more significant as the size of the cluster increases. For example, computational speed of the MLFF is about 40,000 times that of DFT for $Cu_{500}$, as shown in Figure S1 in the supporting information (SI). Therefore, our MLFF can well balance computational accuracy and simulation speed and is competent for searching copper cluster structures.

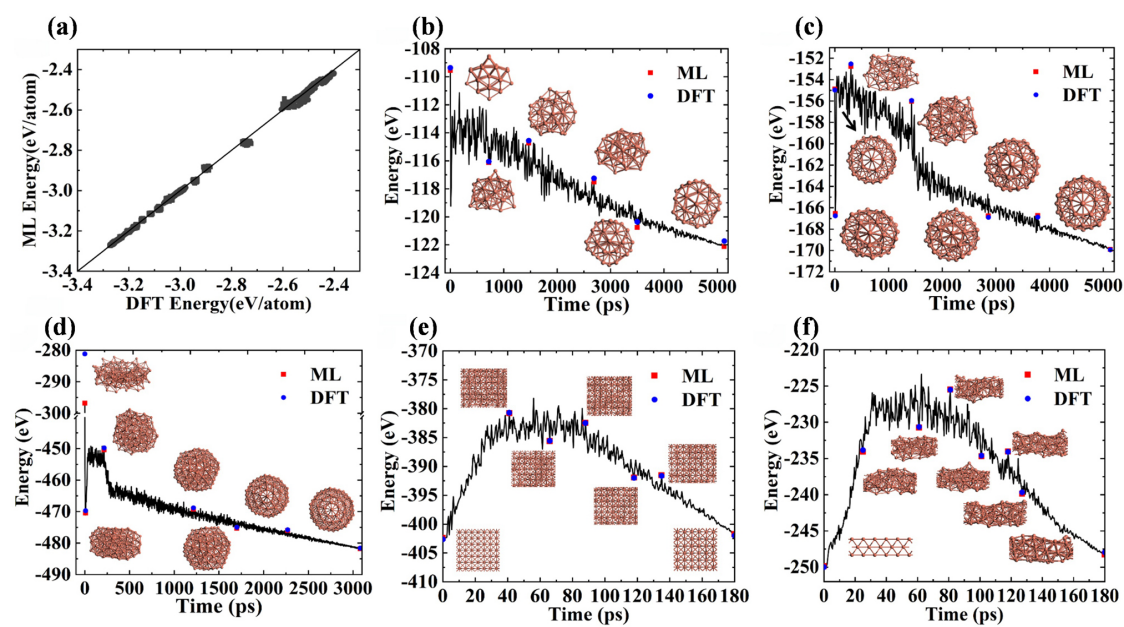

**Figure 2.** (a) Energy comparison between DFT and ML of Cu clusters in the range of $14 \leq N \leq 254$ which are not included in the dataset. The solid line indicates the condition

that ML energy equals DFT energy. The evolution of energy and structures during simulated annealing process for (b) $Cu_{41}$, (c) $Cu_{55}$, (d) $Cu_{147}$, (e) Cu bulk and (f) Cu(001) surface. Insets are randomly selected structures during annealing with energies calculated by both ML (red squares) and DFT (blue dots).

## 2.2 Ground state structures and their stability

With the robust MLFF, simulated annealing was performed for $Cu_N$ clusters ($7 \leq N \leq 80$, $136 \leq N \leq 153$ and $303 \leq N \leq 315$) to search the ground state structures. After annealing, the low energy structures were fully optimized with DFT. The ground state structures for clusters with $N \leq 60$ are shown in Figure S2 in the SI. For all these clusters, the low-energy structures obtained by simulated annealing were fully compared with that predicted previously base on DFT calculations and deep learning [45, 47, 49, 52, 57]. For $Cu_N$ clusters (N = 21, 26, 27, 30, 32, 33, 34, 36, 37, 43, 44, 50), the low-energy structures obtained here are lower in energy than those reported previously[57], as shown in Figure 3. The ground state structures of other clusters obtained here agree well with previous results. Some results align with previous findings, while others surpass prior outcomes, validating the reliability of our MLFF and the effectiveness of our method in searching for cluster structures.

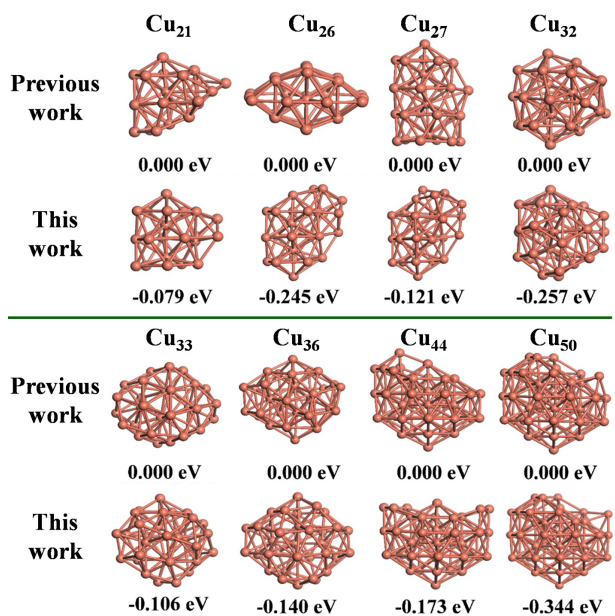

**Figure 3.** Comparison of selected copper clusters obtained in this work and previous work[57].

The relative stability of Cu clusters was revealed by calculating the second-order energy differences $\Delta_2 E$, defined as,

$$\Delta_2 E(N) = E(N+1) + E(N-1) - 2E(N) \quad (1),$$

where, E(N+1), E(N-1) and E(N) are total energies of Cu clusters with N+1, N-1 and N atoms, respectively. As shown in Figure 4a, the positions of the high peaks are the sizes of magic clusters with high relative stability. For small clusters (N < 40), the second-order energy difference exhibits significant even-odd oscillation, as an even number of electrons can ensure that there are no unpaired electrons in the system, thereby reducing the energy. According to electronic counting rules (Jellium model), the size sequence of the magic number cluster of Cu clusters should be 2, 8, 18, 20, 34, 40, 58, 68, 70, 92, 106, 112, 138 and so on. Here, we assume that a copper atom only provides one valence electron. As shown in Figure 4a, in the size range of N < 40, all the electronic magic number clusters show high stability. Unlike electronic magic numbers, the core of geometric magic numbers is the perfect packing of atoms in three-dimensional space, pursuing high symmetry, high coordination number, and low surface energy. According to the icosahedral structure series, the size sequence of the magic number cluster of Cu clusters should be 13, 55, 147, 309 and so on. However, $Cu_{13}$ shows a very low stability. Therefore, at small sizes (N < 40), the electronic counting rule plays a dominant role in the stability of clusters.

For medium sized clusters (40 ≤ N ≤ 80), the phenomenon of odd even oscillations becomes blurred and $Cu_{55}$ with an odd number of electrons exhibits high stability. $Cu_{55}$ is a geometric magic number cluster, so the geometric magic rule also play an important role in the stability. The relative stability of $Cu_{40}$ is not outstanding, even though it is an electronic magic number cluster. $Cu_{58}$, $Cu_{68}$ and $Cu_{70}$ which are electronic magic number clusters, all show high stability. Therefore, for medium sizes (40 ≤ N ≤ 80), both the electronic counting rule and the geometric magic rule contribute.

For large size clusters (N > 80), only $Cu_{147}$ and $Cu_{309}$, which are geometric magic number clusters, show outstanding stability. In contrast, Cu138, an electronic magic number cluster, has a low stability. Therefore, for large size (N > 80), The geometric magic rule plays a dominant role in the stability of clusters.

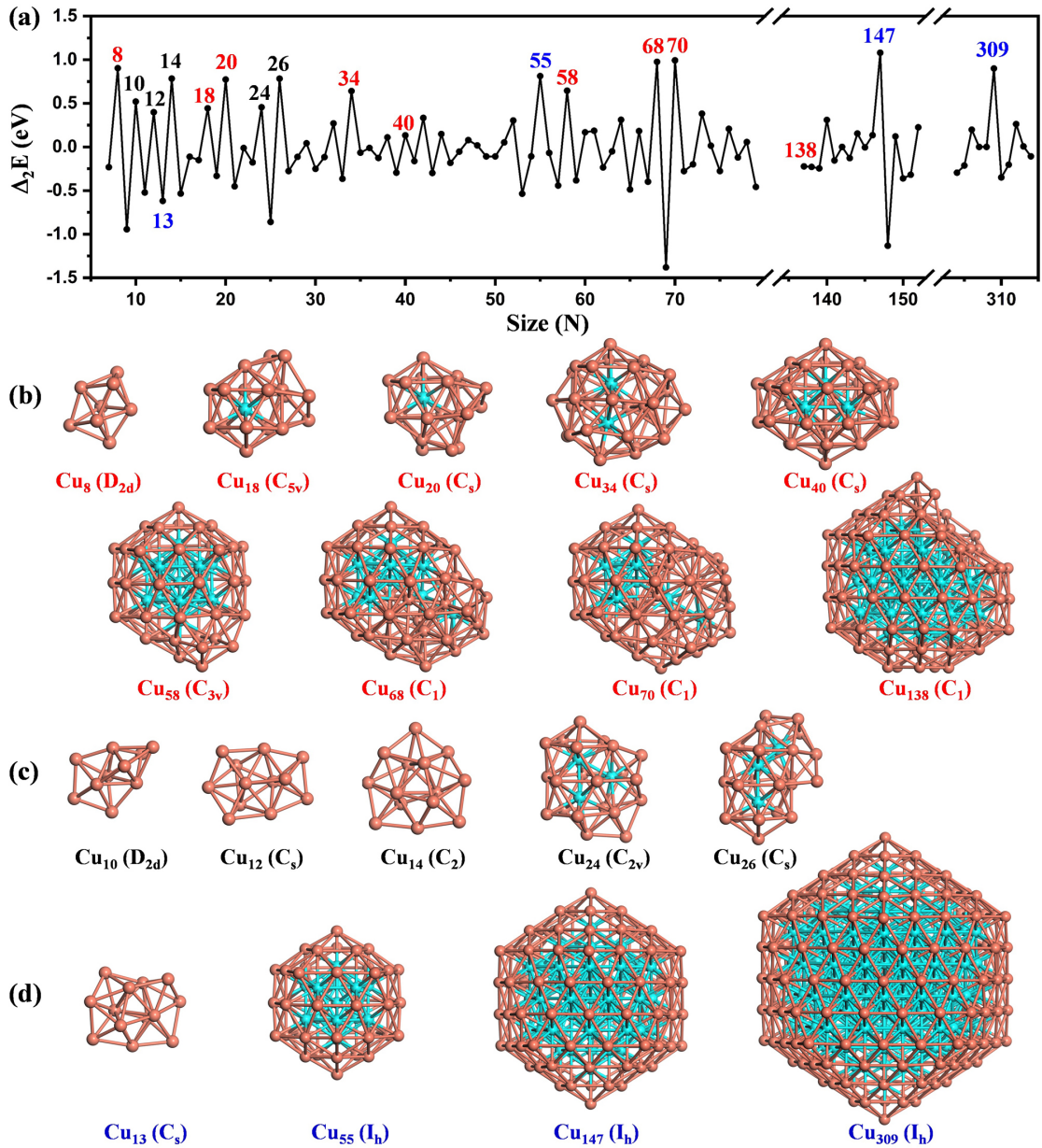

**Figure 4.** (a) Second-order energy differences of Cu$_N$ (7 ≤ N ≤ 315). The red numbers represent electronic magic number sequences, while the blue numbers represent geometric magic number sequences. Ground state structures of (b) cluster with a closed-shell electronic structure, (c) selective clusters with high stability and (d) geometric magic-number clusters. The symmetry groups of the clusters are given in parentheses. Cyan balls represent 12-coordinated copper atoms. Brown balls represent copper atoms with coordination numbers lower than 12.

The ground-state structures of all the highly stable clusters, electronic magic number clusters and geometric magic number clusters are shown in Figure 4b-d. Cu$_8$ shows a high stability due to the high symmetry (D$_{2d}$) and a closed electronic shell. Cu$_{18}$ with a symmetry of C$_{5v}$ is the smallest cluster containing a regular icosahedron and a 12-coordinated Cu atom. Cu$_{20}$ can be seen as adding two Cu atoms on Cu$_{18}$. Cu$_{34}$ obeying the electronic counting rule shows high stability and can be viewed as adding more Cu atoms around the icosahedron. $Cu_{10}$, $Cu_{12}$, $Cu_{14}$, $Cu_{24}$ and $Cu_{26}$ are all very stable due to the lack of unpaired electrons. Moreover, high symmetry also contributes to the stability. For example, the $Cu_{10}$ has a high symmetry of $D_{2d}$, which is higher than its neighbors, $Cu_9$ ($C_{2v}$) and $Cu_{11}$ ($C_2$), as shown in Figure S2 in the SI. $Cu_{26}$ has a symmetry of $C_s$ while $Cu_{25}$ and $Cu_{27}$ are asymmetric ($C_1$) (Figure S2). Thirteen is the minimum size that can form an icosahedral cluster. However, the ground-state structure of $Cu_{13}$ is a structure with a low symmetry of $C_s$. The $Cu_{13}$ cluster with a regular icosahedral structure is 0.97 eV higher in energy than that of the ground state structure, which again proving that for small clusters the electronic counting rule dominates instead of the geometric magic rule.

When N reaches 55, a perfectly nested icosahedron structure with $I_h$ symmetry first forms. Its interior is a small regular icosahedron composed of thirteen 12-coordinated Cu atoms, and its exterior is a large regular icosahedron composed of 42 atoms. $Cu_{58}$, $Cu_{68}$ and $Cu_{70}$ can be viewed as adding more and more Cu atoms on $Cu_{55}$. Because the number of electrons follows the rules of electronic counting, they also show high stability even if the structure is not a perfect icosahedron. The structure of $Cu_{40}$ is far away from a perfect icosahedron, so its relative stability is not very high even though the number of electrons follows the rules of electronic counting.

The ground-state structure of $Cu_{138}$ is an incomplete icosahedral structure with no symmetry. Cu138 shows low stability even though it is an electronic magic number cluster. In contrast, $Cu_{147}$ and $Cu_{309}$ with a complete icosahedral structure and high symmetry ($I_h$) show very high stability, proving that for large clusters the geometric magic rule dominates.

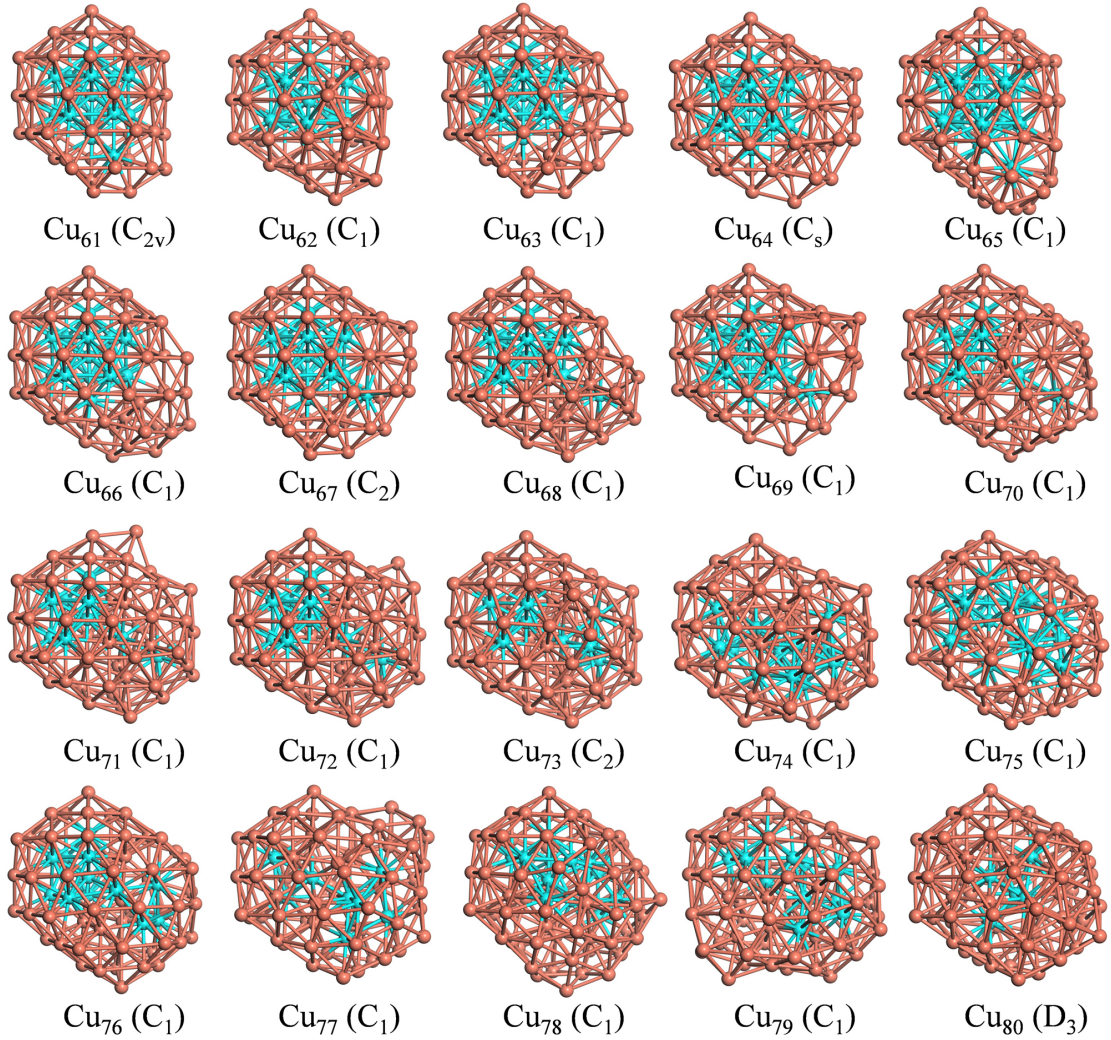

**Figure 5.** Ground-state structures of Cu clusters $Cu_N$ ($61 \leq N \leq 80$) obtained by simulated annealing with MLFF. The symmetry groups of the clusters are given in parentheses. Cyan balls represent 12-coordinated copper atoms. Brown balls represent copper atoms with coordination numbers lower than 12.

In general, Cu clusters tend to form densely packed structures and as the cluster size increases the number of 12-coordinated Cu atoms also increases. The local environment of each 12-coordinated Cu atom is an icosahedron, which is completely different from FCC Cu crystal, where the local environment of each Cu atom is a cuboctahedron. As the cluster size increases, Cu clusters tend to form a core-shell structure nested with icosahedrons. The ground-state structures of Cu clusters $Cu_N$ in the range of $61 \leq N \leq 80$ obtained by simulated annealing with MLFF are shown in Figure 5. Basically, in this size range, the structures of Cu clusters can be viewed as adding Cu atoms on $Cu_{55}$. However, during the continuous evolution of Cu clusters, some distortion was introduced to the $Cu_{55}$ as the core of the cluster, especially when the size is far away from 55, such as $Cu_{75}$ to $Cu_{80}$. Because in the range of medium sizes ($40 \leq N \leq 80$), both the electronic counting rule and the geometric magic rule contribute.

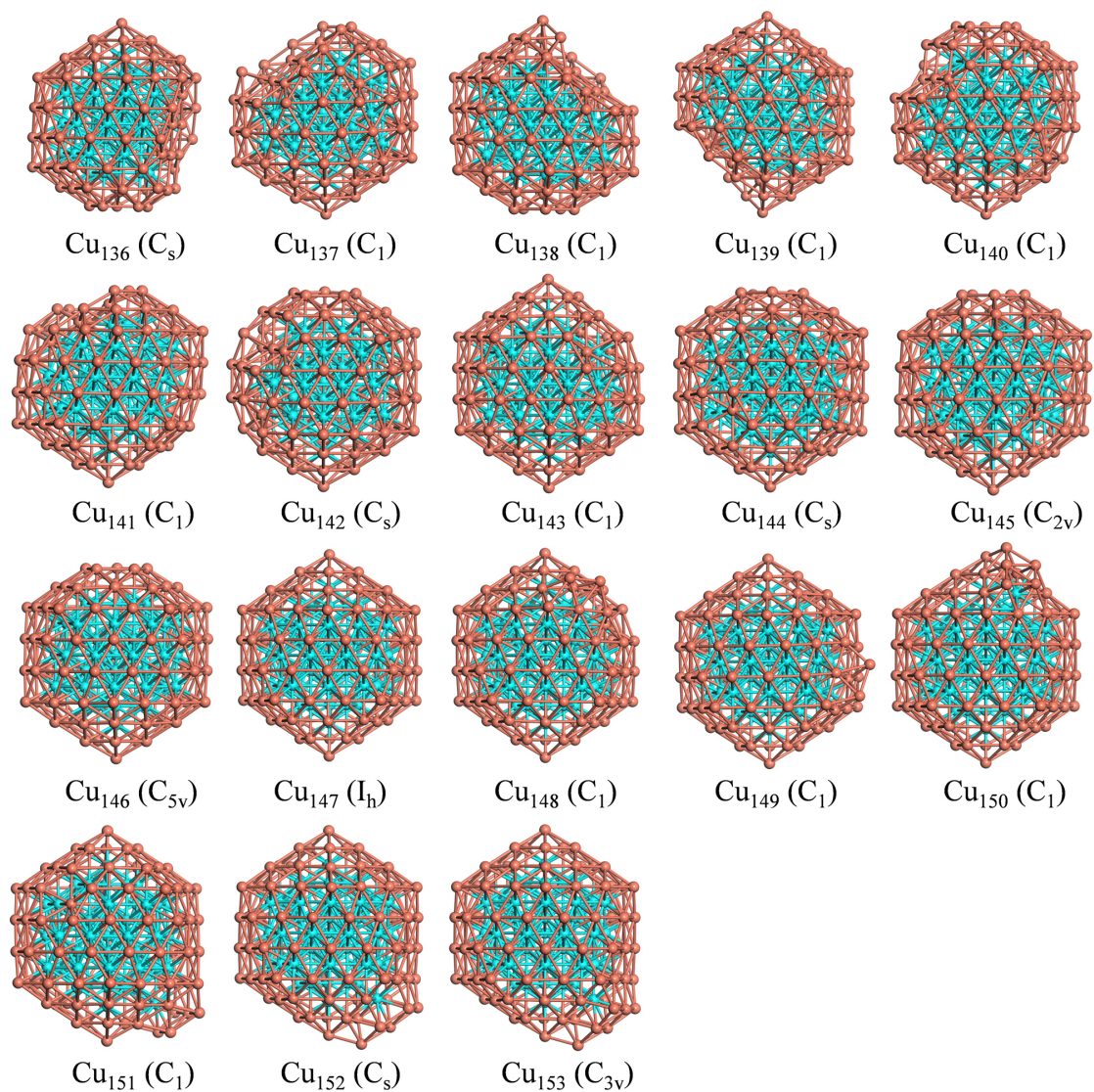

**Figure 6.** Ground-state structures of copper clusters $Cu_N$ ($136 \leq N \leq 153$) obtained by simulated annealing with MLFF. The symmetry groups of the clusters are given in parentheses. Cyan balls represent 12-coordinated copper atoms. Brown balls represent copper atoms with coordination numbers lower than 12.

As the cluster size further increases, the evolution behavior of clusters based on the icosahedral core-shell structure becomes more apparent, as shown in Figure 6. From $Cu_{136}$ to $Cu_{147}$, the third icosahedral shell was gradually completed. From $Cu_{147}$ to $Cu_{153}$, the cluster structure gradually moves towards the fourth icosahedral shell by adding atoms on top of $Cu_{147}$. The icosahedral shells in the core are very stable during the structure evolution. The continuous evolution of Cu clusters around the next geometric magic number cluster ($Cu_{309}$) with the complete fourth icosahedral shell can be seen in Figure S3, which further confirms that the geometric magic number is the only criterion determining cluster stability at large sizes.

## 2.3 From cluster to nanocrystal

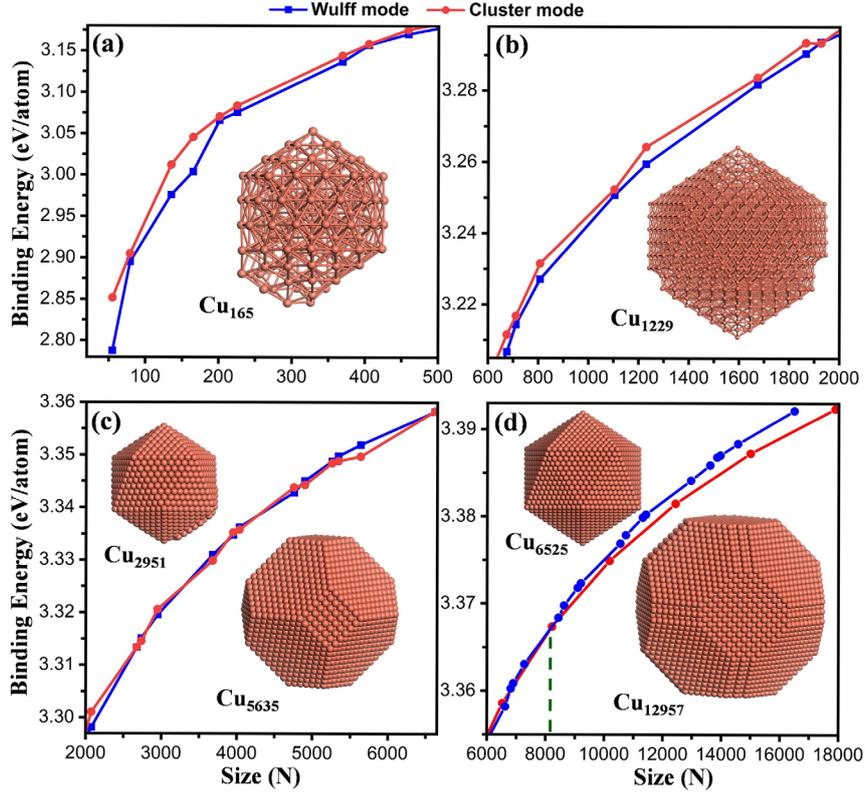

**Figure 7.** Binding energy as a function of cluster size for Cu clusters based on icosahedron structures (Cluster mode) and Wulff structures (Wulff mode). Insets are some typical structures of Cu clusters.

Based on the discussion above, as the number of atoms increases, the cluster will evolve according to the structure of an icosahedron, gradually increasing the number of shell layers of the regular icosahedron, which is referred as Cluster mode. When the size is large enough, the cluster will transform into nanocrystal with crystalline structural characteristics. The geometric structure of nanocrystal can be determined through Wulff construction. Therefore, to search the critical size for the transition from cluster to nanocrystal, we constructed Cu clusters in both Cluster mode and Wulff mode with different sizes ranging from 55 to 6619. The sizes of clusters were chosen according to the Wulff structure so that perfect regular icosahedrons cannot be formed for the clusters in Cluster mode. Clusters in Wulff mode were fully relaxed by MLFF. Clusters in Cluster mode were obtained via mild simulated annealing starting from an incomplete icosahedral structure followed by fully relaxation. Binding energy $E_b$ was calculated to judge the stability, which is defined as,

$$E_b = (N \times E_{Cu} - E_{cluster})/N \quad (2),$$

where, N is the number of atoms in the cluster. $E_{Cu}$ is the energy of an isolated Cu atom. $E_{cluster}$ is the total energy of the cluster. Binding energy as a function of cluster size is shown in Figure 7a-c. When N < 2000, clusters in Cluster mode always have higher binding energies than that of clusters in Wulff mode, even though a perfect icosahedral closed shell layer is not formed (Figure 7a, b). At the range of 2000 ≤ N ≤ 6619, clusters in the two modes have similar binding energies considering that no perfect icosahedral shell can form (Figure 7c).

To clearly see the boundary between Cluster mode and Wulff mode, clusters with perfect icosahedral closed shells and cluster with perfect Wulff constructions were built in the size range of $55 \leq N \leq 17885$. Binding energy as a function of cluster size is shown in Figure 7d. The intersection of the binding energies of the two modes can be clearly seen at about N = 8000. When N > 8000, clusters in Wulff mode always have higher binding energies than that of clusters in Cluster mode, and the binding energy difference always increases with the size increasing. Therefore, when the cluster size exceeds 8000 atoms (about 6 nm in diameter), the cluster will transform into nanocrystal with a crystalline structure.

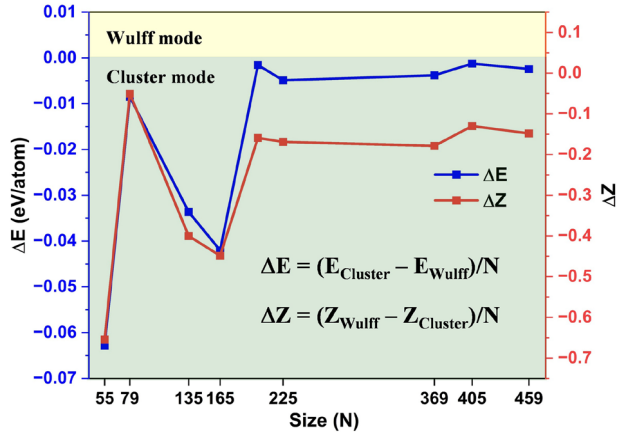

**Figure 8.** The energy difference ΔE and coordination number difference ΔZ of clusters in Cluster mode and Wulff mode as a function of cluster size. In the yellow region, the Wulff mode is stable and in the green region, the Cluster mode is stable.

The reason why the clusters based on icosahedron is so stable is that it has 5 rotational symmetry axes. This symmetry is forbidden in infinitely extended crystals according to the crystallographic restriction theorem, but in finite clusters, it is the perfect way to achieve the densest packing. The icosahedral structure achieves the densest packing (coordination number up to 12) of spheres at a local scale, maximizing interatomic interactions and minimizing surface energy. To verify this point, we further investigated the coordination number of Cu atoms in the clusters under two modes. The coordination number difference ΔZ was calculated to measure the degree of dense stacking, which is defined as,

$$\Delta Z = (Z_{Wulff} - Z_{Cluster})/N \quad (3),$$

where, $Z_{Wulff}$ and $Z_{Cluster}$ are the sum of coordination numbers of all Cu atoms in the cluster in Wulff mode and Cluster mode, respectively. N is the number of atoms in the clusters. Meanwhile, the total energy difference ΔE between the two modes of clusters was also calculated to directly measure the relative stability, which is defined as,

$$\Delta E = (E_{Cluster} - E_{Wulff})/N \quad (4),$$

where, $E_{Cluster}$ and $E_{Wulff}$ are the total energies of the clusters in Cluster mode and Wulff mode, respectively. N is the number of atoms in the clusters. As shown in Figure 8, all the clusters (N ≤ 459) fall into the Cluster mode region because ΔE is negative. ΔZ is always negative, indicating that clusters in Cluster mode are more compact than that in Wulff mode. Importantly, ΔZ and ΔE have the same trend of change. Therefore, the tendency of copper atoms to form structures as densely packed as possible is fundamental to the evolution of copper cluster structures. However, as the cluster size increases, the proportion of low-coordinated Cu atoms at surfaces becomes smaller and smaller, and the more energetically favorable FCC close-packing mode will gradually gain energy advantage.

## 3. Conclusions

In summary, a MLFF applicable to all sizes of Cu clusters was proposed based on equivariant neural networks. Combining this MLFF and simulated annealing, the low energy structures of Cu clusters $Cu_N$ in a wide range (7 ≤ N ≤ 17885) was investigated to get insight into the structural evolution of Cu clusters with size. For small $Cu_N$ clusters (N < 40), electron counting rules play a major role in cluster stability. For medium size clusters (40 ≤ N ≤ 80), the electron counting rule and the geometric magic number rule work together. For large clusters (N > 80), geometric magic number rules play a dominant role in cluster stability. As the cluster size increases, Cu clusters gradually evolve into core-shell structures based on icosahedrons because Cu atoms prefer to form structures as densely packed as possible. As the cluster size increases, the proportion of surface atoms decreases and the Wulff constructed Cu clusters with FCC close-packing mode are gradually energetically preferred, which implies a transition from cluster to bulk phase. The critical size for the transition was calculated to be around 8000 atoms (6 nm in diameter). Our work provides a deep insight into the evolution of Cu clusters and find out the critical size for the transition from Cu cluster to nanocrystal which has long been a great challenge. Our work paves the way for following studies on other clusters.

## 4. Methods

All the first principle calculations were performed by the Vienna ab initio simulation program (VASP)[58]. The electron-ion interaction was described by the projector augmented wave (PAW)[59], and for the exchange-correlation functional, we used the Perdew–Burke–Ernzerhof (PBE) functional within the generalized gradient approximation (GGA)[60]. It was tested that a kinetic energy cutoff of 450 eV for the plane-wave basis group is enough. Convergence criteria of $10^{-5}$ eV and 0.02 eV/Å for the energy and force on each atom was adopted. The Γ-point in the first Brillouin zone is used for calculations of clusters. To avoid fictitious interactions between clusters due to periodic condition, a vacuum of at least 15 Å in all three directions were used.

The MLFF was trained by the Allegro[61] equivariant neural network. Molecular dynamics simulated annealing, performed by the LAMMPS code[62], was used to search low energy structures of Cu clusters. NVT ensemble was used for the simulated annealing. First, the copper clusters were rapidly heated to melt, then slowly cooled to room temperature. The annealing time is greater than 2 ns, depending on the size of the clusters.

## Supporting Information

Supporting Information is available from the Wiley Online Library or from the author.

## Acknowledgments

The authors acknowledge the financial support provided by the National Natural Science Foundation of China (Grant No. 12374174, 12374253), National Key Research and Development Program of China (2024YFE0213500), National Foreign Expert Project (D20240220, D20240213) and R&D project of Joint Funds of Liaoning Province (2023JH2101800038).

## Conflict of Interest

The authors declare no conflict of interest.

## Data Availability Statement

The data that supports the findings of this study are available from the corresponding author upon reasonable request.

## Keywords

# Supporting Information

# From cluster to nanocrystal: the continuous evolution and critical size of copper clusters revealed by machine learning

Hongsheng Liu[1], Luneng Zhao[1], Yaning Li[1], Yuan Chang[1], Shi Qiu[1], Xiao Wang[2,3], Junfeng Gao[1,2]*, Feng Ding[2]

4. Key Laboratory of Materials Modification by Laser, Ion and Electron Beams (Dalian University of Technology), Ministry of Education, School of Physics, Dalian 116024, China
5. Suzhou Laboratory, Suzhou 215123, China
6. Shenzhen Institutes of Advanced Technology, Chinese Academy of Sciences

Corresponding author: (gaojf@dlut.edu.cn)

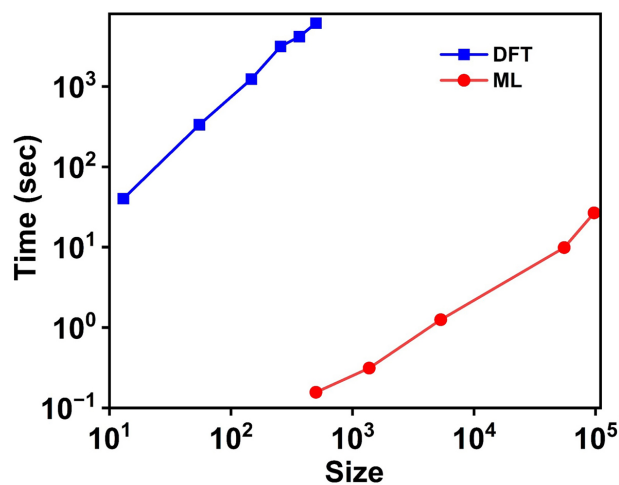

**Figure S1.** The relationship between the time required to calculate a single point energy of a cluster and the size of the cluster.

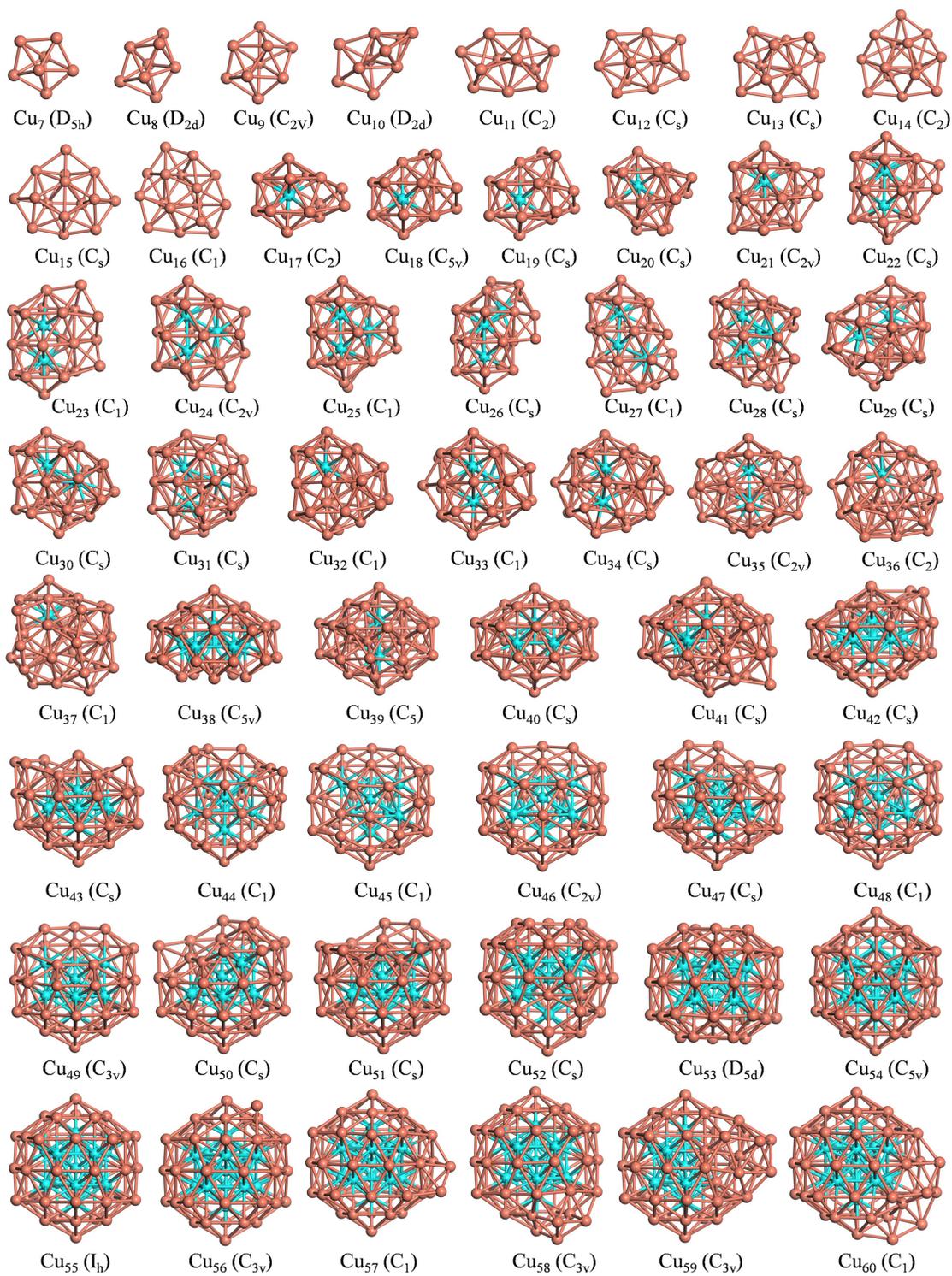

**Figure S2.** Low-energy structure of copper clusters $Cu_N$ ($7 \leq N \leq 60$) obtained by simulated annealing with MLFF. The symmetry groups of the clusters are given in parentheses. Cyan balls represent 12-coordinated copper atoms. Brown balls represent copper atoms with coordination numbers lower than 12.

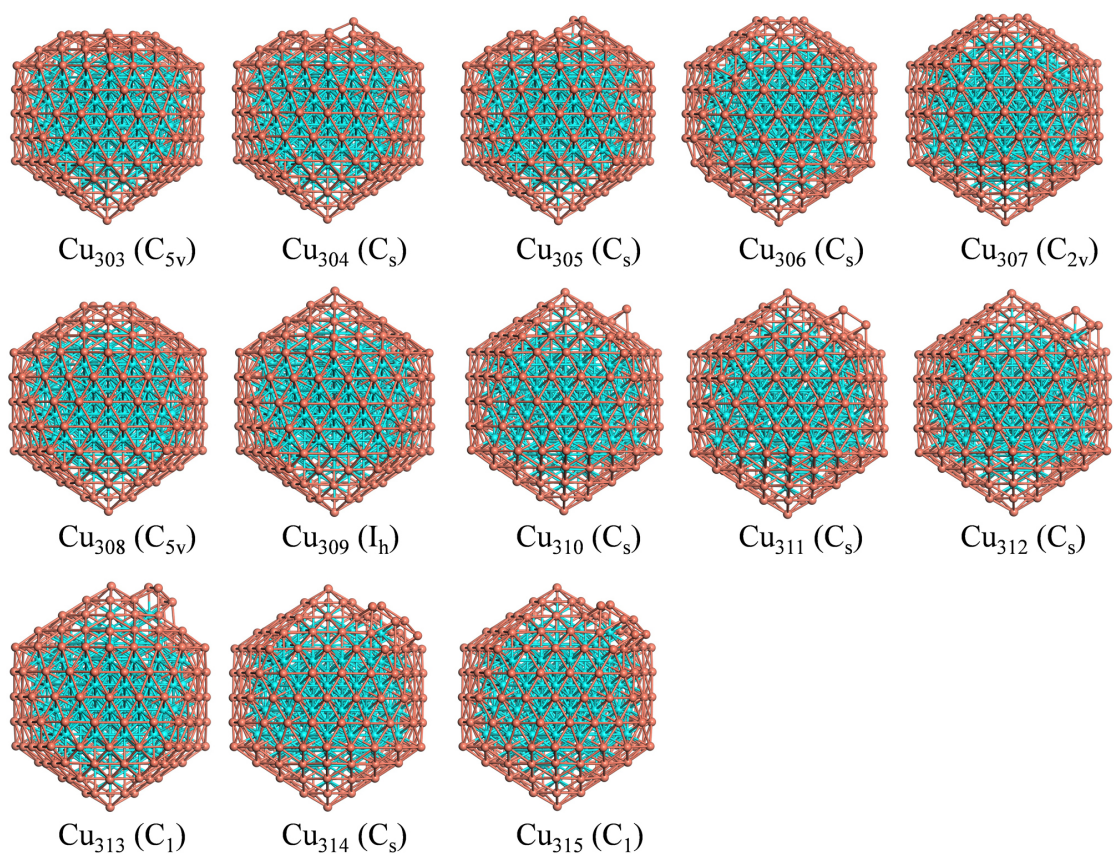

**Figure S3.** Low-energy structure of copper clusters $Cu_N$ ($303 \leq N \leq 315$) obtained by simulated annealing with MLFF. The symmetry groups of the clusters are given in parentheses. Cyan balls represent 12-coordinated copper atoms. Brown balls represent copper atoms with coordination numbers lower than 12.